# The original nucleation and migration of the basal/prismatic interfaces in Mg single crystals

Qun Zu[1], Xiaozhi Tang[1], Shuang Xu[2], Yafang Guo[1]

**Abstract:** The formation of basal/prismatic (BP) interfaces accompanying with the nucleation and growth of a reoriented crystal in Mg single-crystals under *c*-axis tension is investigated by molecular dynamics simulations. The BP interfaces nucleate by shuffling mechanism via local rearrangements of atoms. Both two-layer disconnections and one-layer disconnections contribute to the migration of BP interfaces. In a three-dimensional view, the BP interfaces relatively tend to migrate towards the $[1\bar{2}10]$ direction rather than the $[\bar{1}010]/[0001]$ direction since the misfit disconnection or misfit dislocation caused by the accumulation of mismatch along the $[\bar{1}010]/[0001]$ direction impedes the disconnection movement. The BP interfaces can transform to the $\{10\bar{1}2\}$ twin boundary (TB) and vice versa. While the process from BP interface to TB is described as the linear pile-up of interface disconnections, the versa transformation is proposed as the upright pile-up process. Both BP transformation and $\{10\bar{1}2\}$ twinning can efficiently accommodate the strain along the *c*-axis, and the conjugate BP interfaces and $\{10\bar{1}2\}$ TBs account for the large deviations of twin interfaces from the $\{10\bar{1}2\}$ twin plane.

## 1. Introduction

Deformation twinning plays a crucial role in the plastic deformation of hexagonal close-packed (hcp) metals, for it changes the crystal orientation and simultaneously accommodates the deformation of transverse and longitudinal directions. Due to the pure shear mechanism of twinning, it is suggested that twin boundary (TB) should coincide with the twinning plane, while the interface defects can compensate the small angles deviation of the TB from the twinning plane. However, the recent transmission electron microscopy (TEM) and high-resolution

Yafang Guo: yfguo@bjtu.edu.cn

[1] Institute of Engineering Mechanics, Beijing Jiaotong University, Beijing 100044, China
[2] School of Science, Wuhan University of Technology, Wuhan 430070, China

TEM (HRTEM) images showed that TB deviated significantly from the $\{10\bar{1}2\}$ twinning plane in hcp metals [1-3Zhang2012, Tu2013a, lizhang]. In the experiment of Mg single crystal under the $[1\bar{1}00]$-axis compression and [0001]-axis tension, Liu *et al*. [4,5Liu2014, 2015] observed the twinning-like lattice reorientation without a crystallographic twinning plane. The boundary between the parent lattice and the reorientation lattice is composed predominantly of semicoherent basal/prismatic (BP) interface instead of the $\{10\bar{1}2\}$ TB. Also, Tu *et al*. [6TU2015] identified the irregular-shaped $\{10\bar{1}2\}$ TB in Mg, Ti and Co by TEM and HRTEM, which was closely related to the interface defects and BP interfaces. The BP transformation is widely observed as a significant deformation mechanism in Mg, and it can give persuasive explanation for the TB deviation.

Actually, BP interfaces appeared in several atomistic simulations of the initial twin nucleation and growth [7-10wang2009-5521, wang2009-903，guo2010, Qi2011], but had seldom been particularly mentioned or studied. In our previous simulations of Mg single crystal under *c*-axis tension, BP interfaces were evidently involved with the $\{10\bar{1}2\}$ twinning [9,10guo2010, Qi2011]. BP interfaces were also visible in the tensile simulation of $[11\bar{2}0]$-textured nanocrystalline Mg [11Kim2010] and Zr [12Lu2015], though they were not put forward. In 2013, molecular dynamics (MD) simulations and density functional theory (DFT) calculations demonstrated the nucleation of twins in hcp structure via a pure-shuffle mechanism, while the initial interfaces of the nucleus consist of BP interfaces [13wang2013]. Subsequently, BP interfaces were observed to nucleate in bicrystals with pre-existing $\{10\bar{1}2\}$ TB or BP interface [4，14-17Liu2014NATURE, Ostapovets2014abc,barrett2014a]. In 2014, the BP interface was described as coherent terraces after relaxation, while two types of disconnection dipoles and a misfit dislocation arrange periodically to remove the attendant long-range elastic strains and maximize local coherency. These disconnections promote the migration of the BP segments with the twin boundary [18Barrett2014b]. Moreover, the conjugate PB interfaces and TBs were observed accompanying the growth of a prefabricated twin nucleus [19Xu2013]. Atomistic simulation results indicated that the combined mobility of $\{10\bar{1}2\}$ TB and BP interface controlled the overall kinetics of twin propagation due to their much lower energy.

In our molecular dynamics simulations of the deformation of Mg single crystals with different orientations, BP interfaces have also been observed accompanying with the crystalline reorientation, while BP always connects the coherent TB [20zu2016]. In this paper, we will give further insight on the original nucleation and growth of BP interface in Mg single crystals. The results will be compared with those obtained in previous works using bicrystal or prefabricated nucleus models.

## 2. Simulation model and method

The Mg single crystal column with square cross section is used for studying the crystalline reorientation and BP transformation under the *c*-axis tension. Molecular dynamics method using the LAMMPS code [21Plimpton1995] and EAM potential developed by Liu *et al*. [22liu1997] are applied. We orientate the $[1\bar{2}10]$ direction as the *x*-axis, the [0001] direction as the *y*-axis, and the $[\bar{1}010]$ direction as the *z*-axis in the original model. Free boundary conditions are applied in *x*- and *z*- directions; fixed displacement boundary condition is assigned to the *y*-direction. The dimension of Mg single crystal is about $13\times26\times13$ nm$^3$ in this work, while other sizes of square and circle single crystal columns have also been examined in previous work [20zu2016]. The simulation is performed in constant NVT ensemble with a velocity-Verlet integrator and the temperature is controlled at 5 K by the Nosé–Hoover thermostat. The Atomeye software [23Li2003] is used for visualizing the evolution of the atomistic structures.

In the simulation, the perfect crystal is relaxed for 30 ps at zero force to minimize the potential energy at first. After that, an uniaxial loading is exerted along the *y*-direction by applying a constant strain of about $5\times10^7$ s$^{-1}$ on the 1.0-nm-thick top layer, while the 1.0-nm-thick bottom layer is fixed. The simulations are carried out until the maximal strain reached about 16%. At each loading step, the configuration of the atoms, the atomic energy and stress conditions are recorded for further analysis.

## 3. Results and discussions
### 3.1. Crystalline reorientation under *c*-axis tension

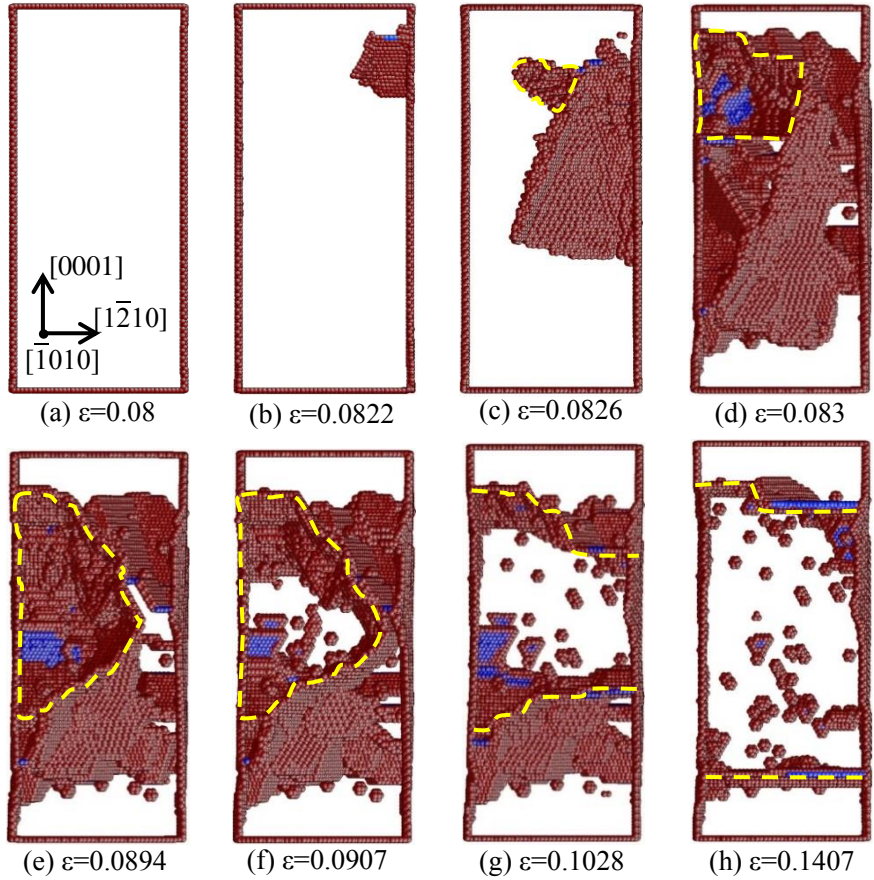

Fig. 1. Plastic deformation evolution of Mg single crystal column under *c*-axis tension viewed along [$\bar{1}010$] direction. (a) Atomistic configurations without hcp atoms in the elastic deformation stage. (b) Partial pyramidal dislocation nucleates in the initial plastic stage. (c-h) The nucleation and growth of the reoriented crystal (in yellow frame).

The plastic deformation evolution under *c*-axis tension of Mg single crystal is shown in Fig. 1, respectively corresponding with the strains of 0.0822, 0.0826, 0.083, 0.0894, 0.0907, 0.1028 and 0.1407. For clarity, the perfect hcp atoms are made invisible, and atoms on stacking fault and other defects such as surfaces and dislocation cores are displayed in blue and red, separately. After the single crystal column elastically deforms up to the yield strain of 0.0822, the plastic phase starts, which is characterized by the nucleation of pyramidal dislocations at the free surface in Fig. 1(b). Subsequently, the reoriented crystal marked by yellow frame nucleates at the surface where the partial pyramidal dislocation locates, and then grows as shown in Figs. 1(c-f). The process is accompanied by the sharp and continuous decrease of the stress. Along with the pyramidal slips, the crystal reorientation dominates the whole deformation process of the single crystal column. At the strain of 0.1028, the reoriented crystal runs through the cross section of the sample shown in Fig. 1(g), and at the strain of 0.1407 it almost fills up the whole single crystal in Fig. 1(h), which separately correspond the fluctuation of the plastic flow stress and the hardening of the single crystal sample. The visualized interface structure and exhaustive formation mechanism of this reoriented crystal are illustrated as follows. In the experimental investigation of the deformation in submicron-sized single crystal Mg, a similar growth process of reoriented crystal with an apparently straight boundary running across the gauge part of the sample was observed, creating an interface without crystallographic twinning plane between the new grain and the matrix [4,5Liu2014, 2015].

## 3.2. Nucleation mechanism of the reoriented crystal

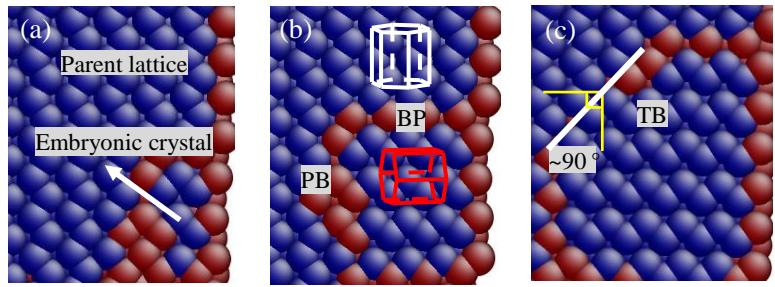

Fig. 2. Nucleation of the reoriented crystal viewed along $[1\bar{2}10]$ direction. The BP/PB boundary between the embryonic crystal and the patent lattice places a basal and prismatic plane face-to-face.

The magnified figures of the nucleation process of reoriented crystal viewed along the $[1\bar{2}10]$ direction at different time steps ($\varepsilon$=0.0824) are displayed in Fig. 2. The embryonic crystal nucleates originally via the local rearrangements of atoms in Fig. 2(a). In Fig. 2(b), the reoriented crystal grows through the BP/PB transformations, indicating the reorientation angle of 90° relative to the parent lattice. Then, $\{10\bar{1}2\}$ TB follows to act as the boundary during the growth of the embryonic crystal as shown in Fig. 2(c). The detailed analysis of the transformation from BP interfaces to TB will be discussed in section 3.5. In summary, the nucleation mechanism of reoriented crystal under *c*-axis tension is described as the process of BP/PB transformation by atomic shuffle.

## 3.3. Structure of the BP interface

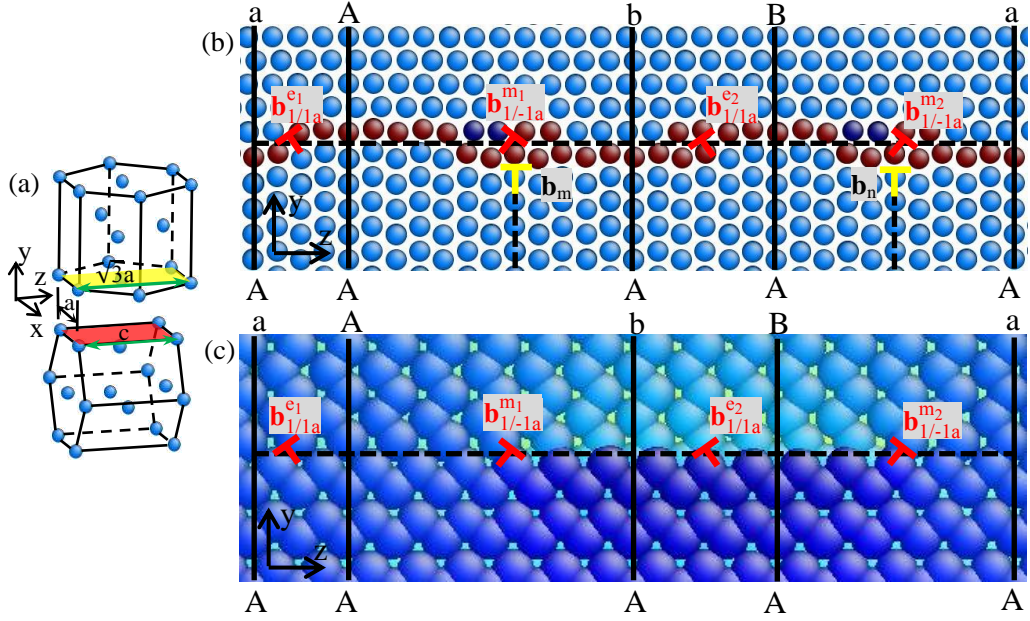

Fig. 3. Atomic structure of the unstressed BP/PB interface. (a) Atomic structure of basal and prismatic unit cells. (b) Interface defects on BP/PB interface identified by CNA. (c) Interfacial structure colored according to *x*-coordinate. The red symbol represents the misfit disconnection, and the yellow symbol indicates the misfit dislocation. The atoms align in layers along $[\bar{1}010]$ direction (upper lattice) and [0001] direction (lower lattice) periodically with the sequence of …AaBbAaBb… and …ABAB…, respectively.

In order to give further studies on the growth and migration of the BP interface, the atomic structure of an unstressed BP interface is presented in Fig. 3. The basal and prismatic unit cells placed face-to-face are shown in Fig. 3(a). It indicates that no mismatch exists in *x*-direction for the upper and lower units. However, the unit length of the upper lattice in *z*-direction ($\sqrt{3}a$) is different from that of the lower lattice (c). Due to the mismatch in *z*-direction, the atoms at the interface adjust their positions to give regions of good and bad registry. Therefore, the BP interface shows a serrated appearance with the disconnections arraying alternatively in Fig. 3(b). Respectively, the atoms in upper and lower lattices align periodically along the *z*-direction with the sequence of …AaBbAaBb… and …ABAB…. The corresponding a-layer in upper lattice and A-layer in lower lattice are chosen to be the reference planes as the a-A-layer. After approximate 15 (upper lattice)-16 (lower lattice) periodic distances along the *z*-direction, the reference planes will almost correspond to each other again, with a small mismatch of $15\sqrt{3}-16\kappa$ ($\kappa=c/a$). Through this pseudo-periodic distance, the A-layer in lower lattice orderly corresponds to a-, A-, b-, B-, and a- layers in upper lattice as shown in Fig. 3(b). Four misfit disconnections $\mathbf{b}^{e(m)_l}_{i/(-)jk}$ or two misfit dislocations $\mathbf{b}_p$ and $\mathbf{b}_q$, exist to accommodate the mismatch along the *z*-direction and displacements along *x*- and *y*- directions. Here *i*, *j* are the numbers of the inter-planar steps in the matrix and reoriented lattices, respectively. "-" is downward disconnection for distinguishing the upper disconnection. *k* indicates the stacking sequence position of {0001} planes on which the disconnection acts. *e* shorts for edge dislocation, *m* shorts for mixed dislocation, and *l* represents the kind of edge (mixed) dislocations. *p* and *q* are used to identify different misfit dislocations.

In order to identify the disconnection type, the same interfacial structure is also colored according to *x*-coordinate in Fig. 3(c). The obvious displacements along *x*-direction are emerged on dark blue atoms. Combined with Fig. 3(b), it can be found that the maximum displacement of *x*-component with a/2 has been completed between A-b-layer and A-B-layer. Thus, the disconnection in this region should be edge type. This means that the edge and mixed disconnections arrange alternately on BP interface. Based on the geometrical analysis and simulation of atomic configuration, the four disconnections are expressed as $\mathbf{b}_{1/1a}^{e_1}$, $\mathbf{b}_{1/-1a}^{m_1}$, $\mathbf{b}_{1/1a}^{e_2}$, and $\mathbf{b}_{1/-1a}^{m_2}$. If the small mismatch of $15\sqrt{3}$-$16\kappa$ along z-direction is ignored, then:

$$\mathbf{b}_{1/1a}^{e_1} = a \begin{pmatrix} 0 \\ \dfrac{\sqrt{3}-\kappa}{2} \\ \dfrac{\sqrt{3}}{6} \end{pmatrix}, \quad \mathbf{b}_{1/-1a}^{m_1} = a \begin{pmatrix} \dfrac{1}{2} \\ \dfrac{\kappa-\sqrt{3}}{2} \\ \dfrac{\sqrt{3}}{3} \end{pmatrix}, \quad \mathbf{b}_{1/1a}^{e_2} = a \begin{pmatrix} 0 \\ \dfrac{\sqrt{3}-\kappa}{2} \\ \dfrac{\sqrt{3}}{6} \end{pmatrix}, \quad \mathbf{b}_{1/-1a}^{m_2} = a \begin{pmatrix} -\dfrac{1}{2} \\ \dfrac{\kappa-\sqrt{3}}{2} \\ \dfrac{\sqrt{3}}{3} \end{pmatrix} \quad (1)$$

Here $\kappa$ is *c/a* ratio, and *a* represents the lattice constant. The Burgers vectors of two edge disconnections are same, while the *x*-components in two mixed disconnections are opposite.

According to the atomic structure of unstressed BP interface, one extra atomic plane emerges in lower lattice compared to the upper crystal in every half pseudo-periodic distance. Thus in a full pseudo-periodic distance, two misfit dislocations are represented by $\mathbf{b}_p$ and $\mathbf{b}_q$, which can be visualized by the accumulation of misfit disconnections with the Burger vector of:

$$\mathbf{b}_p = \mathbf{b}_{1/1a}^{e_1} + \mathbf{b}_{1/-1a}^{m_1} = a \begin{pmatrix} \dfrac{1}{2} \\ 0 \\ \dfrac{\sqrt{3}}{2} \end{pmatrix}, \quad \mathbf{b}_q = \mathbf{b}_{1/1a}^{e_2} + \mathbf{b}_{1/-1a}^{m_2} = a \begin{pmatrix} -\dfrac{1}{2} \\ 0 \\ \dfrac{\sqrt{3}}{2} \end{pmatrix} \quad (2)$$

Barrett et al. [18Barrett2014b] have previously described the BP boundary as semicoherent as well. However, the interfacial characters are slightly different from my analysis, *i.e.* the Burgers vectors and the array of interfacial disconnections. In their paper, the interfacial disconnections exhibited the array of …two mixed characters, two edge characters... with the Burgers vectors of:

$$\mathbf{b}_{mixed} = a \begin{pmatrix} \frac{1}{2} \\ \frac{\sqrt{3}-\gamma}{2} \\ \frac{1}{\sqrt{3}} \end{pmatrix}, \quad \mathbf{b}_{edge} = a \begin{pmatrix} 0 \\ \frac{\sqrt{3}-\gamma}{2} \\ \frac{16\gamma-15\sqrt{3}}{2} + \frac{1}{\sqrt{3}} \end{pmatrix}, \quad \mathbf{b}_{misfit} = a \begin{pmatrix} \frac{1}{2} \\ 0 \\ \frac{-\sqrt{3}}{2} \end{pmatrix} \quad (\gamma = c/a) \quad (3)$$

As we know, it is pseudo-periodic along $[\bar{1}010]/[0001]$ direction. Normally, the screw components should be accompanied in pairs with opposite vectors. However, the opposite screw component was not shown in their work. Furthermore, if the two edge disconnections arrange adjacently as with their simulation, the misfit dislocation in this half pseudo-periodic distance should not have screw component. Therefore, the Burgers vectors of disconnections in my work are more reasonable.

### 3.4. Migration mechanism of BP interface

The migration of BP interface is mediated by the nucleation and movement of disconnections at the BP interface as shown in Fig. 4. Two types of interfacial disconnections are observed, which are indicated as the two-layer step in Fig. 4(a) and one-layer step in Fig. 4(b). In Fig. 4(a1), the junction of a partial pyramidal dislocation (PD) and the BP interface acts as the source of a two-layer disconnection, and a $\mathbf{b}^e_{2/-2}$ step appears when it is viewed along the *x*-direction. In Fig. 4(a2), the two-layer disconnection moves along the *z*-direction, which causes the BP interface migrates upwards (*y*-direction). Subsequently, a new two-layer disconnection forms at the junction of PD and BP interface in Fig. 4(a3), causing the BP moves continuously. Due to the interaction of PD and BP interface, the atoms near the crossing need to accommodate the relatively large and quick deformation. Therefore, the two-layer disconnection contributes to the migration of BP interface along the *y*-direction, as $\mathbf{b}^e_{2/-2}$ described:

$$\mathbf{b}^{\mathrm{e}}_{2/-2} = a \begin{pmatrix} 0 \\ \kappa - \sqrt{3} \\ 0 \end{pmatrix} \qquad (4)$$

Liu *et al*. [liu 2014] also observed the two-layer disconnection in a bicrystal model under a compressive stress perpendicular to the BP interface plane. Ostapovets *et al*. [14Ostapovets2014a] stated the glide mechanism with one-layer and two-layer disconnections on BP interface, yet, the nucleation process of two-layer disconnections was not provided.

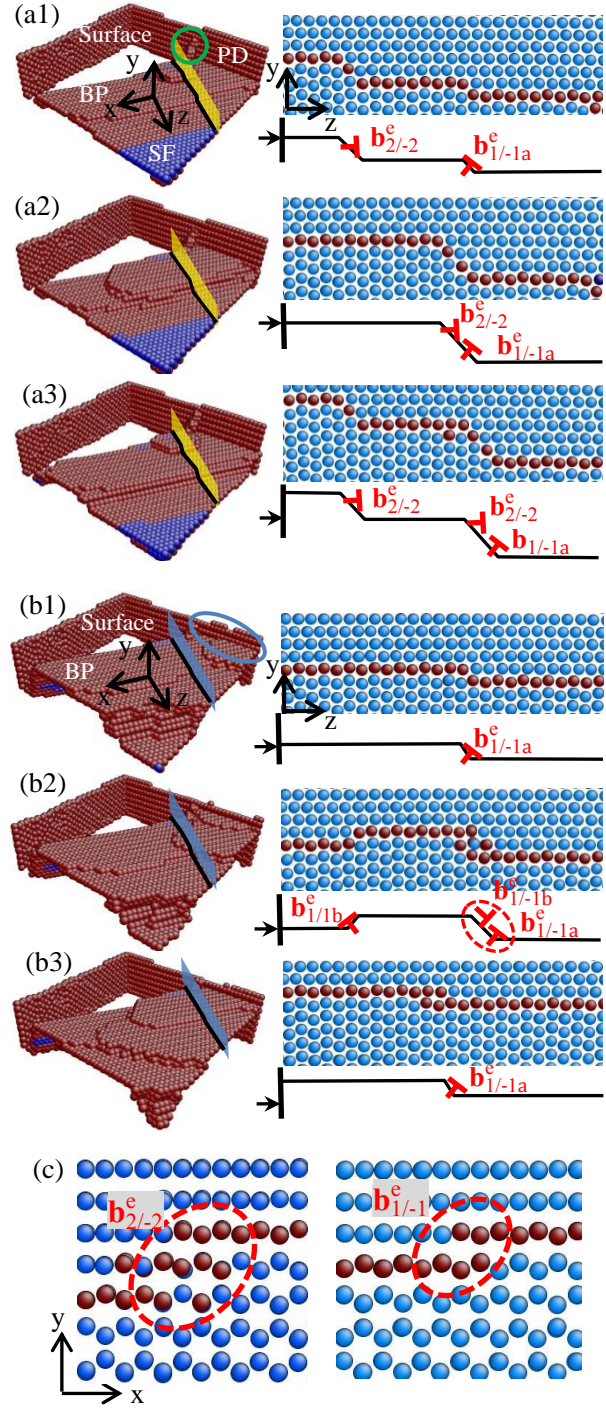

Fig. 4. Migration behavior of BP interface with (a) two-layer disconnection and (b) one-layer disconnection. The junction of partial pyramidal dislocation and BP interface is the source of two-layer disconnection, and that of free surface and BP interface is the origin of one-layer disconnection. (c) The atomic interface configuration viewed along z-direction. The upper and lower regions belong to the parent and reoriented lattices, respectively.

Also, the migration of BP interface is fulfilled by the nucleation of $\mathbf{b}^e_{1/-1b}$ and $\mathbf{b}^e_{1/1b}$ disconnections, as shown in Fig. 4(b). In Fig. 4(b1), new disconnection dipoles nucleate near the junction (blue ellipse) of free surface and BP interface, and next, move on the BP interface (Fig. 4(b2)). Then, the BP interface migrates along the y-direction by continuous nucleation of one-layer disconnections as shown in Fig. 4(b3). Moreover, it is also noticed that the two-layer disconnection ($\mathbf{b}^e_{2/-2}$) can be easily transformed from double one-layer disconnections ($\mathbf{b}^e_{1/-1a}$ and $\mathbf{b}^e_{1/-1b}$). When the later one-layer disconnection ($\mathbf{b}^e_{1/-1b}$) catches up the early one ($\mathbf{b}^e_{1/-1a}$) as shown in Fig. 4(b2), they combines as a two-layer disconnection and move together. The interaction process can be described by:

$$\mathbf{b}^e_{2/-2} = \mathbf{b}^e_{1/-1a} + \mathbf{b}^e_{1/-1b} = a\begin{pmatrix} 0 \\ \dfrac{\kappa - \sqrt{3}}{2} \\ \dfrac{\sqrt{3}}{6} \end{pmatrix} + a\begin{pmatrix} 0 \\ \dfrac{\kappa - \sqrt{3}}{2} \\ -\dfrac{\sqrt{3}}{6} \end{pmatrix} = a\begin{pmatrix} 0 \\ \kappa - \sqrt{3} \\ 0 \end{pmatrix} \quad (5)$$

In Fig. 4(c), the atomic configurations of two-layer disconnection and one-layer disconnection (shown in dash red circles) are presented, showing no atoms relative displacements in x-direction. This means either two-layer disconnection or one-layer disconnection containing no screw components.

It is noticed that the originally nucleation of BP interfaces in a single crystal always occurs at the junction of defects or free surface due to the inhomogeneous local stress, thus the configuration of BP interface is not as regular as the structure in a prefabricated BP interface as shown in Fig. 3(b). It actually shows a terraced appearance, but only contains a few types of disconnections. To say explicitly, the edge type disconnections ($\mathbf{b}^e_{2/-2}$ or $\mathbf{b}^e_{1/-1}$) are popularly observed and responsible for the migration of BP interface due to their wide cores and small Burgers vectors. Moreover, we have obtained an overall view of the BP interfaces migration in Fig. 4. Interestingly, it is found that the BP interfaces relatively tend to migrate towards the x-direction rather than the z-direction. This is because that the misfit disconnection or misfit dislocation caused by the accumulation of mismatch impedes the BP interface movement along the z-direction. When the width of the disconnection dipole reaches a certain value, the nucleation of disconnection dipole via local rearrangements of atoms can occur more easily than the extension of it.

### 3.5. Transformation between BP interface and $\{10\bar{1}2\}$ TB

In MD simulations and experiments, the terraced boundaries consisted of conjugate BP interfaces and $\{10\bar{1}2\}$ TB have been observed in Mg and its alloys [4,6Liu2014,Tu2015]. Thus, it might be a popular behavior that the $\{10\bar{1}2\}$ TB and BP interface transform to each other during the boundary migration. Fig. 5 shows the three-dimension view of the boundary between the parent lattice and the reoriented crystal at the strain of 0.095. The terraced interface is consisted of BP interfaces and $\{10\bar{1}2\}$ TB, together with interface defects on their facets. Both two-layer disconnections and one-layer disconnections exist in the boundary structure. The growth of the reoriented crystal is implemented by the migration of TBs and PB interfaces, accompanying with the transformation between BP interface and $\{10\bar{1}2\}$ TB.

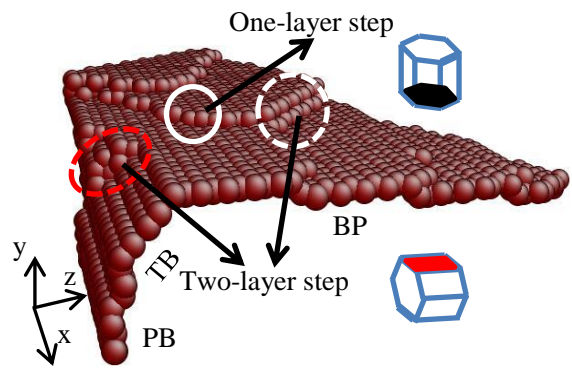

Fig. 5. Three-dimension boundary structure between the parent lattice (upper the boundary) and the reoriented crystal (under the bounday).

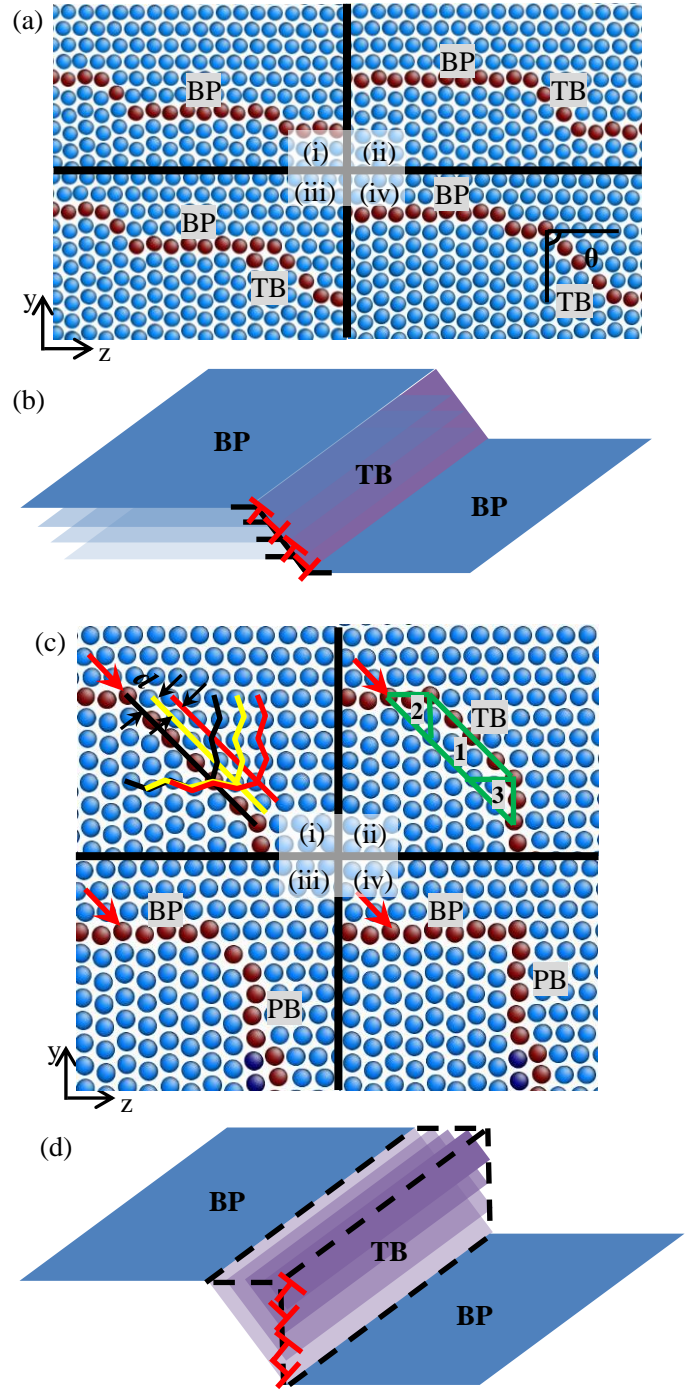

Fig. 6. Transformation process between $\{10\bar{1}2\}$ TB and BP interface. (a) Transformation from BP interface to $\{10\bar{1}2\}$ TB and (b) schematic illustration of this process, fulfilling by the linear pile-up of misfit disconnections. (c) Transformation from $\{10\bar{1}2\}$ TB to BP interfaces and (d) schematic illustration of this process, showing the upright accumulation of disconnections. The changes from the light to dark planes represent the migration processes of BP interface (blue) and $\{10\bar{1}2\}$ TB (purple).

The generation of $\{10\bar{1}2\}$ TB at the BP interface can be described as the linear pile-up of BP interface defects, as shown in Figs. 6(a) and 6(b). Fig. 6(a) indicates that multiple disconnections appear at BP interface to properly accommodate the strain along the loading direction (*y*-direction). As referred to Fig. 4 in the migration mechanism of BP interface, the disconnections can be described by $\mathbf{b}^e_{1/-1a}$, $\mathbf{b}^e_{1/-1b}$ and $\mathbf{b}^e_{2/-2}$. For $\mathbf{b}^e_{1/-1a}$ and $\mathbf{b}^e_{1/-1b}$, the displacement vectors along *z*-direction are opposite, resulting no extra displacement along *z*-direction after superposition. Thus, the movement on the interface can be fulfilled by the accumulation of $\mathbf{b}^e_{1/-1a}$ and $\mathbf{b}^e_{1/-1b}$ followed by the equation (5). The $\{10\bar{1}2\}$ TB is sequentially formed by the linear accumulation of new nucleated disconnections. The simplified schematic of transformation from BP interface to $\{10\bar{1}2\}$ TB is illustrated in Fig. 6(b). The changes from the light blue to dark blue represent the migration process of BP interface accompanying with the nucleation of $\{10\bar{1}2\}$ TB. Since the parent lattice and reoriented crystal are rotated 90° relative to each other and the TB is short, the atoms on TB need to readjust to fit the neighbor atoms. Thus, the angle between the basal planes of parent and reoriented crystals is close to 90° rather than the traditional twin angle of 86.3°. The angle distribution between basal planes of the matrix and embedded twin was investigated by Ostapovets *et al.* [14Ostapovets2014a], also showing the variation of the misorientation near boundaries.

In contrast, the transformation process from $\{10\bar{1}2\}$ TB to BP interfaces has been captured in Fig. 6(c), while the formation of BP interfaces are implemented by the migration of $\{10\bar{1}2\}$ TB. TB migration is carried out by two super $\{10\bar{1}2\}$-layer disconnections, and the glide of twinning dislocation on twin boundary is not observed during this TB migration. The fold lines with different color from black, yellow, to red in Fig. 6(c-i) show the track of the TB migration. Corresponding to the green-1 region in Fig. 6(c-ii), the displacements of $a(\sqrt{3}-\kappa)$ along the *y*-direction and $a(\kappa-\sqrt{3})$ along the *z*-direction are achieved homogeneously. For the junction region of BP/PB interface and $\{10\bar{1}2\}$ TB, the displacements are fulfilled linearly, causing the transformation from TB to BP/PB interface, i.e. linear displacement of $a(\sqrt{3}-\kappa)$ along the *y*-direction for green-2 region and linear displacement of $a(\kappa-\sqrt{3})$ along the *z*-direction for green-3 region. In Fig. 6(c-iv) after the TB disappears, the boundary totally consists of successive BP and PB interfaces. The schematic illustration of transformation from TB to BP/PB interface is shown in Fig. 6(d), assuming that the BP interfaces are fixed. The TB migration is reflected by the transitions from the light to dark purple planes. While the transformation from BP interface to TB can be considered as the linear pile-up of interface disconnections, the transformation from TB to PB can be proposed as the upright pile-up process.

## 4. Discussions

In Mg and its alloys, $\{10\bar{1}2\}$ twins are frequently reported as the extension twins which can accommodate the tensile deformation along the *c*-axis. The lattice strain caused by $\{10\bar{1}2\}$ twinning along the *c*-axis has been calculated as 0.0626 in our previous analysis [24JAP2014]. Meanwhile, the local shear strain is introduced due to the 86.3° misorientation angle of the $\{10\bar{1}2\}$ twinning. The BP transformation is described as the tetragonal compression (under the $[\bar{1}010]$-axis compression condition) rather than the simple shear [5liu2015scripta], thus the lattice strain can be easily calculated as the ideal value of 0.067 via the transformation from basal to prismatic plane. Therefore, both $\{10\bar{1}2\}$ twinning and BP transformation can accommodate the tensile strain along the *c*-axis based on the lattice strain analysis. Furthermore, the BP interface has a much lower interfacial energy (~170 mJ/m$^2$) [19Xu2013] than that of other symmetric tilt grain boundaries (>250 mJ/m$^2$) [25,26wang2010, Ni2015], although it is slightly higher relative to the $\{10\bar{1}2\}$ TB energy (~120 mJ/m$^2$). Thus in our present simulation of Mg single-crystals under *c*-axis tension, the BP transformation coordinating with the $\{10\bar{1}2\}$ twinning accommodate the *c*-axis tensile strain efficiently. The transformations between BP interface and $\{10\bar{1}2\}$ twin boundary (TB) are implemented via the migration of interface disconnections, and the conjugate BP interfaces and $\{10\bar{1}2\}$ TBs account for the large deviations of twin interfaces from the $\{10\bar{1}2\}$ plane. The same phenomena of BP transformation coordinating with $\{10\bar{1}2\}$ twinning have been observed in the deformation of other hcp metals, such as Zr, Co, and Ti [6TU2015]. It indicates that both BP transformation and twinning are important deformation mechanisms for hcp metals.

As the heated debate of the twinning dislocation mechanism and the shuffle mechanism for the $\{10\bar{1}2\}$ twinning [27-29li2009,serra2010,reply2010], we describe the migration of BP interface dominated by the nucleation and glide of interface disconnections via shuffling, just as the conventional glide-shuffle mechanism for twinning. Due to the mismatch of $\{\bar{1}010\}$/$\{0001\}$ boundary, BP interface disconnections are formed with a wide core. The migration of BP interface is implemented by the nucleation and glide of interface disconnections, while different atoms around the disconnections move different distances. This is absolutely different from the traditional glide mechanism of dislocation with a homogeneous shear process. Meanwhile, this shuffling does not mean that the atoms on interfaces could rearrange randomly. It seems that the local atoms rely on the glide of disconnection to carry out the rearrangement.

## 5. Conclusions

Based on the geometrical analysis of atomic configuration, the structure of BP interfaces in Mg is carefully examined. The unstressed BP interface shows a serrated appearance with four disconnections ($\mathbf{b}^{e_1}_{1/1a}$, $\mathbf{b}^{m_1}_{1/-1a}$, $\mathbf{b}^{e_2}_{1/1a}$, and $\mathbf{b}^{m_2}_{1/-1a}$) and two misfit dislocations ($\mathbf{b}_p$, $\mathbf{b}_q$) arraying alternatively to accommodate the mismatch between the corresponding basal/prismatic planes. Moreover, the originally nucleation and migration of the BP interfaces in Mg single crystals under *c*-axis tension are investigated. A new grain forms with the rotation of 90° relative to the parent lattice after the nucleation of BP interfaces via local rearrangements atoms. In the naturally formed BP interfaces in Mg single crystal accompanying with the crystal reorientation, there are only the first type of disconnections ($\mathbf{b}^{e}_{2/-2}$ or $\mathbf{b}^{e}_{1/-1}$) are popularly observed due to their wide cores and small Burgers vectors.

The migration of BP interface is dominated by the nucleation and glide of interface disconnections, including both one-layer ($\mathbf{b}^{e}_{1/-1a}$, $\mathbf{b}^{e}_{1/-1b}$) and two-layer ($\mathbf{b}^{e}_{2/-2}$) disconnections. The junction of free surface and BP interface is seen as the source of one-layer disconnection, and that of partial pyramidal dislocation and BP interface is identified as the source of two-layer disconnection. From an overall view, the BP interfaces always tend to migrate towards the [1$\bar{2}$10] direction rather than the [$\bar{1}$010] direction because the misfit disconnection or misfit dislocation caused by the accumulation of mismatch impedes the BP movement along the [$\bar{1}$010] direction. Meanwhile, {10$\bar{1}$2} TB and BP interface can transform to each other, depending on the pile-up of interface disconnections. The formation of {10$\bar{1}$2} TB is considered as the linear accumulation of BP interface disconnections, while the transformation from {10$\bar{1}$2} TB to PB interface can be proposed as the upright pile-up of disconnections. Both BP transformation and {10$\bar{1}$2} twinning can efficiently accommodate the *c*-axis tensile strain, and the conjugate BP interfaces and {10$\bar{1}$2} TBs accounts for the large deviations of twin interfaces from the $\{10\bar{1}2\}$ plane.

## Acknowledgements